\documentclass[11pt]{article}
\usepackage{latexsym}
\usepackage{amsmath}
\usepackage{tensor}
\usepackage{multibox}
\usepackage{verbatim}
\usepackage{mathrsfs}
\usepackage{amssymb}
\usepackage{epsf}

\DeclareMathAlphabet{\mathpzc}{OT1}{pzc}{m}{it}

 \csname
@addtoreset\endcsname{equation}{section}

\def\prd{\pr \cdot}

\def\gz0{\gamma^{0}}

\def\scs#1{\section{\sc #1}}
\def\scss#1{\subsection{\sc #1}}

\def\a{\alpha}

\def\m{\mu}
\def\n{\nu}

\def\vf{\varphi}

\def\cA{{\cal A}}
\def\cB{{\cal B}}

\def\cF{{\cal F}}

\def\be{\begin{equation}}
\def\ee{\end{equation}}
\def\bea{\begin{eqnarray}}
\def\eea{\end{eqnarray}}
\def\ba{\begin{array}}
\def\ea{\end{array}}
\def\bec{\begin{center}}
\def\ec{\end{center}}
\def\ba{\begin{align}}
\def\ena{\end{align}}

\def\12{\frac{1}{2}}

\def\pr{\partial}
\def\prd{\partial \cdot}

\begin{document}

\begin{flushright}
{\today}
\end{flushright}

\vspace{25pt}

\begin{center}


{\Large\sc Higher Spins and Current Exchanges}\\


\vspace{25pt}
{\sc A.~Sagnotti$^{\; c}$}\\[15pt]

{${}^c$\sl\small
Scuola Normale Superiore and INFN\\
Piazza dei Cavalieri, 7\\I-56126 Pisa \ ITALY \\
e-mail: {\small \it sagnotti@sns.it}}\vspace{10pt}

\vspace{35pt} {\sc\large Abstract}\end{center}

{The simplest higher-spin
interactions involve classical external currents and symmetric tensors $\phi_{\m_1 \ldots \m_s}$, and convey three instructive lessons. The first is a general form of the van Dam-Veltman-Zakharov discontinuity in flat
space for this class of fields. The second is the rationale for its disappearance in (A)dS spaces.
Finally, the third is a glimpse into an option which is commonly overlooked in Field Theory, and which both higher spins and String Theory are confronting us with: one can well allow in the Lagrangians non-local terms that do not spoil the local nature of physical quantities.}

\setcounter{page}{1}

\vskip 40pt

\begin{center}
{\sl Lecture presented at the 9th Hellenic School \\ on Elementary Particle Physics and Gravity, Corfu 2009}
\end{center}


\newpage


\scs{Introduction}\label{sec:intro}

After decades of intense effort, higher-spin gauge fields still define
a vastly uncharted subject \cite{solvay,francia_010}. Even the free theory has been the source of some surprises during the last years, but the frontier has
long been laid by the elusive and contradictory nature of their interactions. Starting from the 1960's a number of authors have considered them on general grounds, realizing that in flat space they are fraught with difficulties that are both conceptual and technical at the same time \cite{diff_HS,otherold,othernew}. These stem, one way or another, from their higher-derivative nature, and point to an inevitably singular behavior if one tries to remove infrared cutoffs, be they mass scales or a cosmological constant. After almost half a century, little is known explicitly about this limiting behavior, but we are well aware of two settings where higher-spin interactions show no signs of inconsistency. The first is String Theory, whose \emph{flat-space} spectra contain infinitely many \emph{massive} higher-spin modes that lie behind its most cherished properties, (planar) duality and a soft ultraviolet behavior. The second is the Vasiliev system \cite{vasiliev}, a set of non-linear equations subsuming the interactions of infinitely many \emph{massless} symmetric fields $\phi_{\m_1 \ldots \m_s}$, with $s=0,1,\ldots$, about \emph{(A)dS backgrounds}. No truly compelling reason exists at present to focus on the \emph{massless} limit for higher-spin fields in flat space, but our experience with low spins strongly suggests that masses should ultimately arise from the breaking of gauge symmetries. Hence, it is difficult to resist the feeling that higher spins will play a role in an eventual, more satisfactory formulation of String Theory, or that they will help to unveil possible generalizations.

This lecture is meant to summarize some results concerning the simplest interactions of massless higher-spin fields, those induced by external currents. In Electromagnetism they simply subsume, in flat space, Coulomb's law for static charges or its Lienard-Wiechert extensions, but the generalization to higher spins is to embody a surprising feature that emerged long ago for spin two. Even after letting the mass tend to zero, massive gravitational exchanges yield in fact results that are strikingly different from those obtained directly for massless fields. In contrast, as was observed in \cite{porrati}, the current exchanges of a spin-2 field of mass $M$ in a dS background of radius $L$ display a smooth behavior, in such a way that the flat-space discontinuity results formally from the distinct limiting behaviors of analytic, and actually rational functions of $(ML)^2$, at the origin and at infinity.
In Sections \ref{sec:vdvz} and \ref{sec:AdS} I thus review how this van Dam-Veltman-Zakharov discontinuity \cite{vdvz} and the corresponding analytic behavior in an (A)dS background were extended in \cite{fms1} and \cite{fms2} to the whole class of symmetric $\phi_{\m_1 \ldots \m_s}$ tensors. The curved-space analysis is considerably more complicated, but has the virtue of displaying clearly the reason behind the rational functions of $(ML)^2$ and thus, a fortiori, behind the vDVZ discontinuity of these free theories. Finally, there is a third aspect of the work, also reviewed in Section \ref{sec:vdvz}, whose general lesson pervades \cite{fms1,fms2} but was perhaps not fully stressed there. The local Lagrangians that lie behind these current exchange amplitudes were obtained in \cite{fsold,st,fms1,fms2} enlarging Fronsdal's original setting \cite{fronsdal} via the minimal set of fields that make an unconstrained gauge symmetry possible\footnote{A non-minimal local unconstrained form was previously obtained in \cite{bpt}.}. If these additional fields are eliminated via their field equations, as was recently described in \cite{francia_010}, specific \emph{non-local}
Lagrangian forms for the geometric equations of \cite{fsold} obtain, the same that were in fact selected in \cite{fms1} by the condition that their current exchanges coincide with those of the \emph{local} theory. One is thus confronted with a simple instance of a phenomenon that could well play a more prominent role in Field Theory in the future: in theories with a large gauge symmetry the presence of \emph{non-local} terms can be perfectly compatible with the existence of well-defined \emph{local} observables.

\vskip 24pt

\scs{Flat Space exchanges and the vDVZ discontinuity}\label{sec:vdvz}

\scss{The local theory}

The starting point for the massless current exchanges in flat space are the Lagrangian compensator equations of \cite{fsold,st,fms1} in the presence of a \emph{conserved} external current $J$,
\be \cA \ - \ \frac{1}{2} \ \eta \ \cA^{\; \prime} \ + \ \eta^2 \ \cB
= \ J \, , \label{eqaJ} \ee
where
\be \label{tensorA}
\cA \, = \, {\cF} \, - \, 3 \, \pr^{\, 3} \, \alpha \,
\ee
is the minimal local gauge invariant completion of Fronsdal's tensor $\cF$ \cite{fronsdal} for a spin-$s$ field $\vf_{\m_1 \ldots \m_s}$ and $\alpha$
denotes the corresponding spin-$(s-3)$ compensator $\a_{\m_1 \ldots \m_{s-3}}$. As in the original papers, I am resorting to a compact index-free notation that is explained in detail in \cite{cfms} together with its extension to mixed-symmetry fields.

In order to obtain the current exchanges, eq.~\eqref{eqaJ} is to be first combined with its traces in order to solve for ${\cal A}$, which is finally to be folded into the external conserved current $J$. In this fashion one is led to
\be \label{fmsfin}
{\cal E}_f (s,d) \, \equiv \, J \, {\cal A} \, = \, \sum_n \,  {\Gamma\left(3-\frac{d}{2}-s\right) \, s! \over n!\, (s-2n)!\, 2^{2n}\, \Gamma\left(3-\frac{d}{2}-s+n\right)}\ J^{[n]} \; J^{[n]} \, ,
\ee
for the $d$-dimensional massless exchanges, where $J^{[n]}$ denotes the $n$-th trace of the external current. The Euler $\Gamma$ functions encode the dependence of the current exchanges on the space-time dimension $d$, which reflects their close link with the actual physical modes. For $s=1$, taking into account current conservation reduces indeed $J_\m \, J^\m$  to $J_i\, J^i$, the sum over the transverse physical components. This type of reduction to the physical modes reflects the unitarity of the theory, and continues to take place for higher spins. For $s \geq 2$, however, it entails the removal of traces, an operation that depends crucially on the dimension $d$ of space time and is the origin of the van Dam-Veltman-Zakharov discontinuity. In fact, for any given spin-$s$ field one can \emph{define} the massive exchanges in $d$ dimensions starting from a massless $(d+1)$-dimensional field with a prescribed harmonic dependence on a circle coordinate. Let me stress that, in this fashion, one is describing nothing more than an ordinary massive field, since all additional $d$-dimensional fields thus introduced disappear after fixing some Stueckelberg symmetries. However, the dependence on $d$ of the flat-space massless exchanges \eqref{fmsfin} makes the results markedly different from their massive counterparts, so that the comparison between ${\cal E}_f (s,d+1)$ and ${\cal E}_f (s,d)$ extends the result of \cite{vdvz} to the whole class of $\vf_{\m_1 \ldots \m_s}$ fields.

\scss{The non-local theory}

The local compensator equations \eqref{eqaJ} were obtained elaborating on a \emph{non-local} extension \cite{fsold} of Fronsdal's theory \cite{fronsdal} that is free of its (double-)trace constraints. In the absence of external currents, the resulting geometric equations can be presented in the suggestive non-Lagrangian forms
\be \frac{1}{\Box^{\, p}} \, \prd  {\cal R}^{\, [p]}{}_{;\, \m_1 \cdots
\m_{2p+1}} \,  = \,  0 \label{oddcurv} \ee
for odd spins $s=2p+1$, and
\be \frac{1}{\Box^{\, p-1}} \ {\cal R}^{\, [p]}{}_{;\, \m_1 \cdots
\m_{2p}} \ =\ 0 \label{evencurv} \,  \ee
for even spins $s=2p$, that are \emph{non local} for $s\geq 3$ but include both the Maxwell equations $\partial_\m F^{\m\n}=0$ and the linearized Einstein equations $R_{\m\n}=0$.

A key issue in \cite{fms1} was the search for an action principle leading to eqs.~\eqref{oddcurv} and \eqref{evencurv} that, in the presence of external currents, would lead precisely to the exchanges of eq.~\eqref{fmsfin}. The Lagrangian originally proposed in \cite{fsold} fulfills only the first of these conditions, while the proper solution, which is in fact \emph{unique}, was originally obtained in \cite{fms1} adapting the non-local construction, insofar as possible, to its local counterpart. The recent analysis of \cite{francia_010} links the local and non-local forms of the theory via the elimination of compensators and Lagrange multipliers.

The non-Lagrangian equations \eqref{oddcurv} and \eqref{evencurv} are the one option involving, for any $s$, the smallest negative power of the D'Alembertian operator, and thus the mildest type of non locality. In the absence of external currents they can be turned into other forms containing higher negative powers, higher-order poles for brevity, but the different choices become inequivalent when external currents are present. And, as we have stated, there is a \emph{unique} non-local Lagrangian giving rise to the current exchanges \eqref{fmsfin}: it is however more complicated than the expression proposed in \cite{fsold}, due to a tail of higher-order poles whose length grows with $s$. This flat-space theory also admits a massive deformation without the need for any additional fields \cite{dario_massive}, whose current exchanges reproduce directly the vDVZ discontinuity. The mere existence of these geometric Lagrangians may be regarded, in my opinion, as a glimpse of a general lesson: theories containing non-local terms that are compatible with, and are actually required by, an extended gauge symmetry can yield proper and well-defined local physical observables.

\vskip 24pt

\scs{(A)dS exchanges and partially massless fields}\label{sec:AdS}


Massive (A)dS exchanges in $d$ dimensions can be built starting again from massless flat-space ones in $(d+1)$ dimensions, and thus from eqs.~\eqref{eqaJ}, via a procedure usually termed ``radial dimensional reduction'' \cite{radialred}, that may be regarded as a variant of the polar decomposition of flat space but entails some novel complications.

The $s=2$ dS amplitude was actually worked out directly in \cite{porrati}, and the result,
\be
{\left( J_{\mu\nu}\right)^2  \ - \ \frac{1}{d-1}\
{(ML)^2-(d-1)\over (ML)^2-(d-2)}\ \left({J^{\;
\prime}}\right)^2} \ , \label{s2ads}
\ee
is an interesting rational function of $(ML)^2$, where $M$ is the mass of the spin-two field and $L$ is the dS radius. The corresponding AdS amplitude can be obtained letting $L^2\to - L^2$. At any rate, eq.~\eqref{s2ads} recovers the massless $d$-dimensional $s=2$ exchange of eq.~\eqref{fmsfin} in the limit $ML \to 0$, and its $(d+1)$-dimensional counterpart, that as we have explained describes the corresponding massive exchange, in the limit $ML \to \infty$. In other words, the $s=2$ vDVZ discontinuity draws its origin from two distinct limits of an analytic function.

The most interesting feature of eq.~\eqref{s2ads} is its pole: it lies precisely where \emph{partial masslessness} \cite{partlymass} presents itself in a dS space for $s=2$. For $(ML)^2=d-2$, in fact, the massive dS theory acquires the peculiar gauge symmetry
\be
{\delta \vf_{\mu\nu} = \nabla_{\mu}\nabla_{\nu} \, \zeta \,+\, \frac{1}{ L^2}\ g_{\mu\nu} \, \zeta} \ , \label{partials2}
\ee
so that a conserved current does not suffice to guarantee the ``partially massless'' gauge invariance of the coupling $\vf_{\m\n} J^{\m\n}$. Rather, under the transformation \eqref{partials2} this term varies proportionally to the trace of $J^{\m\n}$, which is neatly reflected in the nature of the singular term in \eqref{s2ads}. To reiterate, for $s=2$ the pole reflects a clash between conserved currents and the partially massless gauge symmetry emerging at a special value of $(ML)^2$.

For the general (A)dS exchanges, one can continue to work in a gauge where all internal components of the gauge fields vanish, but even currents that are conserved in the $(d+1)$-dimensional flat space and are covariantly conserved in the $d$-dimensional dS space have now radial components. Moreover, after eliminating the compensator $\a$, a task that is much harder to achieve in this case, the resulting expressions are \emph{not} manifestly analytic in $(ML)^2$, which only enters via
\be \nu={d\,+\, 2\, s\, +\, 1\over 4}\, +\, \frac{i}{2}\, \sqrt{(ML)^2-\left(\frac{d}{2}\, +\, s \, -\, {5\over
2}\right)^2}\, . \label{masshell}\ee

The (A)dS exchanges are indeed analytic functions of $(ML)^2$ for any $s$, thanks to the properties of the generalized hypergeometric function ${_3F_2}\big(a,b,c;d,e;z\big)$, that for $|z|<1$ can be defined as
\be
_3F_2\big(a,b,c;d,e;z\big)\, =\, \sum_{n=0}^\infty \frac{\Gamma(a+n)\, \Gamma(b+n)\,
\Gamma(c+n)\, \Gamma(d)\, \Gamma(e)}{\Gamma(d+n)\, \Gamma(e+n)\,\Gamma(a)\, \Gamma(b)\,
\Gamma(c)} \ \frac{z^n}{n!} \, , \label{3F2}
\ee
and can be related to the more familiar ${_2F_1}$ according to
\be
{_3F_2}\big(a,b,c;d,e;z\big)\, = \, \frac{\Gamma(e)}{\Gamma(c)\Gamma(e-c)} \
\int_0^1 dt \; t^{c-1} \, (1-t)^{e-c-1} \; {_2F_1}(a,b;d;tz)
\ee
within regions for the parameters where all these quantities are defined.
To make a long story short, the analytic continuation of $_2F_1$ beyond the disk $|z|<1$ induces that of $_3F_2$, which is the key ingredient to show that the exchanges are indeed analytic in $(ML)^2$.

Aside from the $s=1,2$ cases, the first few (A)dS current exchanges are:
\begin{eqnarray}
& {\bf s=3 :}  \quad  & \left( J_{abc} \right)^2 \ - \
\frac{3}{d+1}\ {(ML)^2-(d+1)\over (ML)^2-d}\ \left({J^{\;
\prime}}{}_{\; a} \right)^2 \\
& {\bf s=4 :}  \quad  & \left( J_{abcd} \right)^2 \ - \
\frac{6}{d+3}\ {(ML)^2-(d+3)\over
(ML)^2-(d+2)}\ \left({J^{\; \prime}}{}_{\; ab} \right)^2 \nonumber \\
& \quad &+ \, \frac{3}{(d+1)(d+3)} \ \frac{(ML)^4 -
4(ML)^2(d+1)+3(d+1)(d+3)}{[(ML)^2-(d+2)][(ML)^2- 3 d]}\ \left({J^{\;
\prime\prime}}\right)^2 \nonumber\\
& {\bf s=5 :}  \quad  & \left( J_{abcde} \right)^2 \ - \
\frac{10}{d+5}\ {(ML)^2-(d+5)\over (ML)^2-(d+4)}\ \left({J^{\;
\prime}}{}_{\; abc} \right)^2 \\
& \quad &+ \, \frac{15}{(d+3)(d+5)} \ \frac{(ML)^4 -
4(ML)^2(d+3)+3(d+3)(d+5)}{[(ML)^2-(d+4)][(ML)^2- 3 (d+2)]}\
\left({J^{\;
\prime\prime}}{}_{\;a}\right)^2 \nonumber\\
& {\bf s=6 :}  \quad  & \left( J_{abcdef} \right)^2 \ - \
\frac{15}{d+7}\ {(ML)^2-(d+7)\over (ML)^2-(d+6)}\ \left({J^{\;
\prime}}^{\; abcd} \right)^2 \\
& \quad &+ \, \frac{45}{(d+5)(d+7)} \ \frac{(ML)^4 -
4(ML)^2(d+5)+3(d+5)(d+7)}{[(ML)^2-(d+6)][(ML)^2- 3 (d+4)]}\
\left({J^{\;
\prime\prime}}{}_{\;ab}\right)^2 \nonumber\\
& &\hspace{-2cm}- \, \frac{15[(ML)^6 -(ML)^4(9d+31)+
23(ML)^2(d+3)(d+5)-15(d+3)(d+5)(d+7)]}{(d+3)(d+5)(d+7)[(ML)^2-(d+6)][(ML)^2-
3 (d+4)][(ML)^2- 5 (d+2)]}\
\left({J^{\;
\prime\prime\prime}}\right)^2 \nonumber\\
 \end{eqnarray}
Notice that their poles lie precisely at the special ``partially massless'' points where, as in eq.~\eqref{partials2}, some gauge parameters are not accompanied by gradients.


\vskip 24pt
\scs{Acknowledgments}


I am very grateful to Dario Francia and Jihad Mourad for extensive
discussions and collaboration on the material discussed in this lecture, and to the Organizers for their kind invitation to lecture in Corfu. The present
research was supported in part by the European ERC Advanced Grant 226455 SUPERFIELDS, by
Scuola Normale Superiore, by INFN and by the MIUR-PRIN contract 2007-5ATT78.

\vskip 36pt


\end{document}